\documentclass[aps,a4paper,10pt,twocolumn,nofootinbib]{revtex4} 
\usepackage[T1]{fontenc}
\usepackage{mathptmx}
\usepackage{datetime}
\usepackage{amsmath}
\usepackage{amsfonts}
\usepackage{amsfonts}
\usepackage{mathrsfs}
\usepackage[mathscr]{euscript}
\usepackage[dvips]{graphicx}
\usepackage{fancyhdr}
\usepackage{colordvi}
\usepackage{hyperref} 
\usepackage{epsfig}
\usepackage{color}
\usepackage{bm}
\usepackage{tikz}
\usetikzlibrary{decorations,decorations.text,arrows.meta}

\def\overstrike#1#2{{\setbox0\hbox{$#2$}\hbox to \wd0{\hss
    $#1$\hss}\kern-\wd0\box0}}

\renewcommand{\Vec}{\bm}

        \DeclareMathOperator{\grad}{\nabla}
        \DeclareMathOperator{\cross}{\times}

\newdateformat{yymmdddate}{\THEYEAR/\twodigit{\THEMONTH}/\twodigit{\THEDAY}}

\newcommand{\XDOI}[1]{\href{http://dx.doi.org/#1}{doi:#1}}
\newcommand{\XARXIV}[1]{\href{http://arxiv.org/abs/#1}{arXiv:#1}}
\newcommand{\XWEB}[1]{\href{#1}{#1}}

\definecolor{XGREEN}{rgb}{0.01, 0.70, 0.01}
\definecolor{XRED}{rgb}{0.70, 0.01, 0.01}
\definecolor{XBLUE}{rgb}{0.01, 0.01, 0.70}

\newcommand{\pForce}{F}              
\newcommand{\pForcev}{\Vec{\pForce}} 

\newcommand{\pSpeed}{v}                  
\newcommand{\pVelocity}{\Vec{\pSpeed}}   

\newcommand{\pPermittivity}{\epsilon}      
\newcommand{\pPermittivityVac}{\epsilon_0} 

\newcommand{\pPermeability}{\mu}           
\newcommand{\pPermeabilityVac}{\mu_0}      

\newcommand{\pXemXelectric}{E}                      
\newcommand{\pXemXelectricv}{\Vec{\pXemXelectric}}  
\newcommand{\pXemXdisplacement}{D}                         
\newcommand{\pXemXdisplacementv}{\Vec{\pXemXdisplacement}} 
\newcommand{\pXemXpolarization}{P}                         
\newcommand{\pXemXpolarizationv}{\Vec{\pXemXpolarization}} 

\newcommand{\pXemXmagnetic}{B}                           
\newcommand{\pXemXmagneticv}{\Vec{\pXemXmagnetic}}       
\newcommand{\pXemXmagstrength}{H}                        
\newcommand{\pXemXmagstrengthv}{\Vec{\pXemXmagstrength}} 
\newcommand{\pXemXmagnetization}{M}                          
\newcommand{\pXemXmagnetizationv}{\Vec{\pXemXmagnetization}} 

\newcommand{\pCharge}{q}                   
\newcommand{\pCurrent}{J}                  
\newcommand{\pCurrentv}{\Vec{\pCurrent}}   

\newcommand{\pEfieldv}{\pXemXelectricv}
\newcommand{\pDfieldv}{\pXemXdisplacementv}
\newcommand{\pPfieldv}{\pXemXpolarizationv}
\newcommand{\pBfieldv}{\pXemXmagneticv}
\newcommand{\pHfieldv}{\pXemXmagstrengthv}
\newcommand{\pMfieldv}{\pXemXmagnetizationv}

\def\pAfieldv{\Vec{A}}
\def\pCharged{\rho}

\def\pScurrentv{\Vec{\sigma}}

\def\DH{(\pDfieldv,\pHfieldv)}
\def\EB{(\pEfieldv,\pBfieldv)}

\def\bulk{\textup{B}}
\def\bound{\textup{b}}
\def\free{\textup{f}}
\def\SC{\textup{S}}
\def\crc{\textup{c}}
\def\vac{\textup{vac}}

\newcommand{\dT}[1]{\dot{#1}}
\newcommand{\norm}[1]{\left\lVert{#1}\right\rVert}
\newcommand{\defn}[1]{\textbf{{#1}}}

\begin{document}
\title{Maxwell's $\DH$ excitation fields: lessons from permanent magnets}
\author{Jonathan Gratus$^{1,2}$}
\homepage[]{https://orcid.org/0000-0003-1597-6084}
\email[\hphantom{.}~]{j.gratus@lancaster.ac.uk}

\author{Paul Kinsler$^{1,2}$}
\homepage[]{https://orcid.org/0000-0001-5744-8146}
\email[\hphantom{.}~]{Dr.Paul.Kinsler@physics.org}

\author{Martin W. McCall$^{3}$}
\homepage[]{https://orcid.org/0000-0003-0643-7169}
\email[\hphantom{.}~]{m.mccall@imperial.ac.uk}

\affiliation{
  $^1$Physics Department,
  Lancaster University,
  Lancaster LA1 4YB, 
  United Kingdom.
}

\affiliation{
    $^2$Cockcroft Institute, 
    Sci-Tech Daresbury, 
    Daresbury WA4 4AD, 
    United Kingdom.
}

\affiliation{
  $^3$Department of Physics,
  Imperial College London,
  London SW7 2AZ,
  United Kingdom.
}


\begin{abstract}


Macroscopic Maxwellian electrodynamics
 consists of four field quantities
 along with electric charges and electric currents.
The fields occur in pairs, 
 the primary ones being 
 the electric and magnetic fields $\EB$, 
 and the other 
 the excitation fields $\DH$.
The link between the two pairs of field is provided
 by constitutive relations, 
 which specify $\DH$ in terms of $\EB$; 
 this last connection 
 enabling Maxwell's (differential) equations 
 to be combined in a way that supports waves.
In this paper we examine the role played by the excitation fields $\DH$, 
 showing that they can be regarded as not having a physical existence, 
 and are merely playing a mathematically convenient role.
This point of view is made particularly relevant 
 when we consider competing constitutive models of permanent magnets, 
 which although having the same measurable magnetic properties,
 have startlingly different behaviours for the 
 magnetic excitation field $\pHfieldv$.

\end{abstract}


\date{\today}
\maketitle
\thispagestyle{fancy}

%

%
\section{Introduction}\label{S-intro}

The role and meaning of the Maxwell excitation fields $\DH$ 
 has been a subject of debate,
 with arguments for and against 
 their independent existence and measurability
 \cite{Heras-2011ajp,Roche-2000ajp,Scheler-P-2015ejp,Roche-1998ejp,Landini-2014piers,Bork-1963ajp}.
However, 
 a recent work has demonstrated
 that in certain physical circumstances,
 such as when a black hole forms then evaporates,
 $\DH$
 \emph{cannot} be guaranteed to be uniquely defined
 \cite{Gratus-KM-2017nocharge}.
This reduces $\DH$ into the role of a gauge field 
 for the current, 
 meaning that there is no role 
 for attempting to measure them.

Here
 we advance the argument
 that $\DH$ are not physical fields
 equal in status to $\EB$,
 and do so in a way intended to be 
 more concrete and 
 comprehensible to a wider audience.
In the textbook by Jackson \cite{Jackson-ClassicalED}
 for example, 
 the author states clearly that $\DH$ are ``derived fields''
 used as a ``convenience'', 
 but does not 
 discuss their measurability.
There is also plenty of valid electromagnetic theory 
 in which all four fields are treated as having 
 the same status, 
 ranging over traditional textbooks
 \cite{RMC,Jackson-ClassicalED},
 pre-metric electrodynamics \cite{HehlObukhov}, 
 and others \cite{Kinsler-FM-2009ejp}.
Generally, 
 these rely on the existence of agreed constitutive relations, 
 although from the pre-metric electrodynamics \cite{HehlObukhov} point of view,
 the constitutive relations include the role of a metric, 
 rather than being only the material responses.
Further, 
 in some aspects of optics and engineering, 
 $\pHfieldv$ is treated as if
 \emph{it} were the fundamental magnetic field
 instead of $\pBfieldv$, 
 although this is based more on a long-standing calculational convenience, 
 rather than the result of taking a deliberate theoretical stance.

Beyond an examination of Maxwell's equations themselves, 
 we consider 
 three possible constitutive models that can 
 be used to describe magnetism:
 bulk magnetic response,
 surface currents, 
 and a distribution of magnetic dipoles.
Although an initial response might be to expect the models to differ, 
 and consider the fact unremarkable, 
 here we use it to drive home the necessarily ambiguous interpretation
 of the $\DH$ fields.

In what follows, 
 we follow an
 undergraduate-level presentation 
 of electromagnetism
 and the role of its excitation fields $\DH$. 
We do this with a particular focus on
 the treatment of magnetization
 using different constitutive models.
In Sec. \ref{S-Maxwell} we start by briefly discussing
 Maxwell's equations and the role of gauge transformations, 
 and
 then in Sec. \ref{S-measurement} consider the role of measurement 
 and electromagnetic constitutive relations.
Next, 
 in Sec. \ref{S-magnets}
 we compare three different models for the constitutive relations
 of a permanent magnet, 
 and show how although each gives the same result
 exterior to the magnet, 
 the specific $\pHfieldv$ required is not the same, 
 and it can even point in opposing directions.
These inconsistencies between models
 and the difficulties that arise with surface properties
 are emphasised by considering 
 infintely large magnetic slabs
 in Sec. \ref{S-paradoxiis},
 after which
 we conclude in Sec. \ref{S-conclusion}.

%
\section{Maxwell's Equations}\label{S-Maxwell}

Maxwell's macroscopic equations for electromagnetism
 \cite{RMC,Jackson-ClassicalED}
 are well known, 
 and in their Heaviside (vectorial) form, 
 with an over-dot denoting a time derivative,
 are:
~
\begin{align}
  \grad \cdot \pBfieldv &= 0, \qquad~
  \grad \cross \pEfieldv + \dT{\pBfieldv}= 0, 
\\
  \grad \cdot \pDfieldv &= \pCharged_{\free}, \qquad
  \grad \cross \pHfieldv - \dT{\pDfieldv}= \pCurrent_{\free}.
\label{eqn-MaxwellStandard}
\end{align}
Here the fields $\pEfieldv$ and $\pBfieldv$ are the fundamental 
 electric and magnetic fields, 
 whereas $\pDfieldv$ and $\pHfieldv$ are their related excitation fields
The excitation field
 $\pDfieldv$ is  called the ``displacement field'', 
 and $\pHfieldv$
 is often\footnote{Further, 
   $\pHfieldv$ might also be called, 
   depending on context, 
   the ``magnetic field intensity''
   or the ``magnetizing field''.
  Sometimes $\pHfieldv$ 
   is even called the ``magnetic field'',
   which is why $\pBfieldv$ also has alternative names, 
   notably ``magnetic flux density'' or ``magnetic induction''.
  This ambiguity is why we prefer the strict naming of ``fields $\EB$''
   and ``excitation fields $\DH$''.}
 known as the ``magnetic field strength''.
Lastly, 
 $\pCharged_{\free}$ and $\pCurrent_{\free}$ are
 the free charge and current densities.

Like the fields $\EB$, 
 the excitation fields $\DH$ 
 only appear
 in Maxwell's equations with a derivative operator
 applied to them.
This means that $\DH$ 
 are invariant under the addition
 of a total derivative,
 i.e.
~
\begin{align}
  \pHfieldv &\rightarrow \pHfieldv + \grad \psi + \dT{\Vec{v}}
\label{eqn-DHgaugeH}
\\
  \pDfieldv &\rightarrow \pDfieldv + \grad \cross  \Vec{v}
.
\label{eqn-DHgaugeD}
\end{align}
 where $\psi$ and $\Vec{v}$ are arbitrary scalar and vector fields. 
As far as Maxwell's equations are concerned
 we may regard the excitation fields $\DH$
 as a potential field for $(\pCharged_{\free}, \pCurrentv_{\free})$. 
This is not dissimilar to the way 
 we can regard quantities $(\phi, \pAfieldv)$
 as potentials for $\EB$,
 where $\EB$ are invariant under the gauge transformation
~
\begin{align}
  \phi &\rightarrow  \phi  + \dT{\varphi}
\\
  \pAfieldv &\rightarrow \pAfieldv + \grad \varphi
.
\label{eqn-Agauge}
\end{align}
where $\varphi$ is an arbitrary scalar field.
It is worth noting that although many
 do not regard the electromagnetic potentials
 $(\phi, \pAfieldv)$
 as being real physical (measurable) quantities, 
 they are often treated as exactly that --
 notably, 
 in quantum electrodynamics \cite{Louden-TQTL}, 
 where it is the electromagnetic potential $\pAfieldv$ that is quantized, 
 not the fields $\pEfieldv$ or $\pBfieldv$.
Of course, 
 in some cases,
 e.g. in the field of quantum optics,
 there can be a preference to instead quantize the dual potential
 \cite{Hillery-M-1984pra,Datta-1984ejp,Quesada-S-2017ol},
 as a technical device to assist in
 the handling of the material response.


One way of avoiding the indeterminacy of $\DH$
 is to choose to start with the microscopic Maxwell's equations.
In this case
 we can express the constitutive relations
 for the material properties
 solely in terms of bound charges $\pCharged_{\bound}$
 and currents $\pCurrentv_{\bound}$.
The physical picture attached to these properties
 is that they are due to charges attached 
 (``bound'') to an atom, 
 molecule, 
 or some other similarly localized object.
Thus if an applied electric field polarizes an atom, 
 displacing its orbiting electrons, 
 the atom is no longer a simple neutral object, 
 but instead appears as a bound electric dipole.
Similarly, 
 an electron orbiting an atom 
 can appear as a bound current loop.
We treat these bound charges and currents 
 $(\pCharged_{\bound}, \pCurrentv_{\bound})$
 separately from the free charges and currents 
 $(\pCharged_{\free}, \pCurrentv_{\free})$.
We have therefore that
\begin{align}
\grad \cdot \pBfieldv& =0, \qquad&
\grad \cross \pEfieldv + \dT{\pBfieldv}  &=0, \qquad
\\
  \pPermittivityVac
  \grad\cdot \pEfieldv 
&=
  \pCharged_{\free}
 + \pCharged_{\bound}
\quad&\textrm{and}\quad
  \frac{1}{\pPermeabilityVac}
  \grad\cross \pBfieldv
 -
  \pPermittivityVac 
  \dT{\pEfieldv}
&=
  \pCurrentv_{\free} + \pCurrentv_{\bound}
.
\label{CP_Max_bdd}
\end{align}
Next, 
 changing to a macroscopic view,
 we can relate these bound charges and currents to 
 polarization fields  $\pPfieldv$ and $\pMfieldv$ 
 in the usual way, 
 i.e.
\begin{align}
  \pCharged_{\bound}
&=
 -\grad\cdot \pPfieldv, 
\qquad&
  \pCurrentv_{\bound}
&=
  \grad\cross\pMfieldv
 +
  \dT{\pPfieldv}, \qquad
\label{CP_def_BoundrJ}
\\
\quad\textrm{with}\quad
  \pDfieldv 
&=
  \pPermittivityVac \pEfieldv
 +
  \pPfieldv, 
\qquad&
  \pHfieldv
&=
  \frac{1}{\pPermeabilityVac} 
  \pBfieldv  
-
  \pMfieldv
.
\label{CP_def_Bound}
\end{align}
We can immediately see that since $\pPfieldv$ and $\pMfieldv$
 only appear when subject to derivative operators, 
 they are not uniquely determined 
 by $\pCharged_{\bound}$ and $\pCurrentv_{\bound}$.
Of course, 
 although the polarisation $ \pPfieldv $ and magnetisation $\pMfieldv$
 have the same gauge freedom as \eqref{eqn-DHgaugeH}, \eqref{eqn-DHgaugeD}, 
\begin{align}
\pMfieldv  \to \pMfieldv  - \grad\psi - \dT{\Vec{v}}
\quad\textrm{and}\quad
 \pPfieldv  \to  \pPfieldv  + \grad\cross\Vec{v}
,
\label{CP_MP_gauge}
\end{align}
the bound charge and current are gauge invariant. 
Thus we can see that the attempt (or choice)
 we make to use $\pPfieldv$ and $\pMfieldv$ 
 instead of $\pCharged_{\bound}$ and $\pCurrentv_{\bound}$
 necessarily introduces an ambiguity
 into our description.
Our subsequent choice of polarization gauge or detailed constitutive model
 is then driven by the form in which we find
 that the constitutive relations are best expressed, 
 and not by any objective physical reasons.
Notably,
 if we assume that the polarization and magnetization
 depend on the $\EB$ fields, 
 e.g. $\pPfieldv = \hat{\pPfieldv}(\pEfieldv)$
 for some function $\hat{\pPfieldv}$,
 then the
 constitutive relations give $\pCharged_{\bound}$ and $\pCurrent_{\bound}$
 in terms of (the derivatives of) 
 $ \pEfieldv $ and $ \pBfieldv $.

%
\section{Measurement of the electromagnetic and excitation fields}\label{S-measurement}

Our ability to directly measure the properties
 of a system 
 is a 
 primary test of whether or not we consider those properties
 real and physical --
 if we cannot directly measure them, 
 it becomes possible to debate 
 whether they have an independent physical existence at all.

The most direct way of measuring $\EB$ 
 is using the Lorentz force law, 
 which lets us calculate the acceleration
 of a known charge $\pCharge$
 due to those $\EB$ electromagnetic fields,
 i.e.
~
\begin{align}
 {\pForcev}
&=
 \pCharge \pEfieldv + \pCharge \pVelocity \cross \pBfieldv
.
\label{eqn-LorentzForceLaw}
\end{align}
Even the generalization of this
 to the case of hybrid electric-magnetic charged dyons
 depends on $\EB$ 
 and not $\DH$ \cite{DosSantos-2015ejp}.
Another point to note that using \eqref{eqn-LorentzForceLaw}
 is really only practical in free space,
 as inside a medium the path of an electron 
 is too dependent on the material's atomic or molecular structure.

Alternatively,
 we might use the Aharonov-Bohm effect
 \cite{Ehrenberg-S-1949prsb,Aharonov-B-1959pr,Matteucci-IB-2003fp,Batelaan-T-2009pt}
 where we track the phase of electrons around a block of material 
 by relating a closed electron path 
 to properties on a surface bounded by that path.
The advantage of this is that while the electrons may pass 
 around the outside of a block of matter,
 the surface $S$ can pass though the medium. 
This enables one to measure the magnetic flux -- 
 i.e. the integrated $\pBfieldv$ --
 inside the medium.
For some surface $S$ whose boundary 
 $\partial S$
 matches the trajectory of a particle with charge $q$, 
 the phase shift $\theta$ induced on that particle 
 is determined by the integral
~
\begin{align}
  \theta
=
  \frac{q}{\hbar}
  \oint_{\partial S} \pAfieldv \cdot dl 
=&
  \frac{q}{\hbar}
  \int_S \pBfieldv \cdot dS
.
\label{eqn-AharBohm}
\end{align}
~
Thus we can directly measure $\pBfieldv$ in a medium, 
 at least as it is averaged over the cross-sectional surface $S$.
A complementary experiment enables us to measure $\pEfieldv$ in the medium
 \cite{Batelaan-T-2009pt}.

In contrast,
 whether $\DH$
 might also be \emph{directly} measurable fields
 in the manner of $\EB$
 is at best uncertain.
Firstly, 
 there is no well established force law involving $\DH$,
 and even those that there are 
 specify the forces on magnetic monopoles
 \cite{Rindler-1989ajp}.
Secondly, 
 no Aharonov-Bohm-type concept exists for $\DH$.
Of course 
 \emph{if} the $\pEfieldv$ or $\pBfieldv$ has been measured 
 at a point in vacuum, 
 we can use the defined vacuum constitutive relations of
 $\pDfieldv=\pPermittivityVac\pEfieldv$
 and
 $\pBfieldv=\pPermeabilityVac\pHfieldv$
 to also determine $\DH$.
However this \emph{inference} of $\DH$
 is entirely dependent on measurements taken of $\EB$, 
 and then converted with the help of the vacuum constitutive relationship.

Indeed, 
 close examination of all the cases we have seen
 where $\DH$ is claimed to be measured
 (e.g. note in particular the careful 
  theoretical arguments in \cite{HehlObukhov}, 
  or the more practical ones in \cite{Scheler-P-2015ejp})
 it turns out that they are not directly measured, 
 but inferred from a measurement of $\EB$ and
 a model of the constitutive relations,
 i.e. the permittivity and permeability is posited
 and the free parameters of the model determined by experiment. 
For example the permeability of iron in a static field
 $\pPermeability_{\textup{iron}}$
 can be determined by measuring
 the effect in the vacuum of an external magnetic field
 when a piece of iron is introduced. 
However,
 the interpretation of this result
 depends on a model of the magnetic constitutive relation, 
 i.e.  $\pBfieldv = \pPermeability_{\textup{iron}} \pHfieldv$. 
It is this model that ties $\pHfieldv$ to $\pBfieldv$,
 and so prevents one performing the 
 gauge transformation \eqref{eqn-DHgaugeH} \eqref{eqn-DHgaugeD} on $\DH$.

As a result, 
 if we are unable or unwilling to assume a constitutive model,
 we are forced to then claim it is not possible to measure $\DH$, 
 even indirectly.
This is important,
 for example when we consider homogenisation procedures
 \cite{HomogMetaMat-2013pnfa}
 which claim to determine effective constitutive relations. 
Therefore we need to identify what properties
 of the effective constitutive relations are being assumed. 
The freedom is even larger when we permit
 the effective constitutive relations
 to have more than just a temporal response, 
 i.e. if they can also be 
 spatially dispersive
 \cite{AgranoGinsberg,Mnasri-KSPR-2018prb,Khrabustovskyi-MPSR-2017arxiv,Belov-MMNSST-2003prb,Agranovich-G-2006pu,Gratus-M-2015jo,Boyd-GKL-2018oe-tbwire,Gratus-KMT-2016njp-stdisp}.

In what follows we will show that various different constitutive models
 for permanent magnets 
 do not and cannot agree on how to infer 
 $\pHfieldv$ from $\pBfieldv$,
 even though each is widely used and uncontroversial.
This demonstrates not only the central role of the constitutive model, 
 but also the lack of any independent meaning 
 in the $\DH$ excitation fields.

%
\section{Permanent Magnets}\label{S-magnets}

The primary reason that a consideration of magnetism 
 exposes the role of constitutive models most clearly
 is due to the absence of magnetic charges.
For the electric response of a material, 
 it is obvious that one can model this by something as straightforward
 as a polarization field induced by bound charges.
Athough one might, 
 as an exercise, 
 model the electric response as loops of magnetic current charges
 (see e.g. \cite{Kinsler-FM-2009ejp}), 
 it would not ordinarily be seen as an approach
 that represents some underlying microscopic physical model.
However, 
 with no magnetic monopoles to work with, 
 it is entirely reasonable to represent a magnetic response 
 as being generated either by loops of electric current
 or by magnetic dipoles \cite{RMC,Bezerra-KCF-2012ejp,Seleznyova-SK-2016ejp}.

There are three models of permanent magnets
 that we wish to consider here.
The first two models
 of the magnetic constitutive relation of a permanent magnets, 
 which we call the \defn{bulk} and \defn{surface-current} models,
 are both valid
 but have contrasting implementations.
A third,
 which we call the \defn{microscopic} model,
 uses \eqref{CP_Max_bdd},
 \eqref{CP_def_BoundrJ}
 to replace the $\pHfieldv$-field with the bound current. 
This model,
 however,
 is less useful as the bound current depends on
 the derivative of the magnetic field $\pBfieldv$.

These models all contain two distinct contributions, 
 the first due to the bulk properties of the material making up the magnet, 
 and its response to the applied magnetic field
 as defined by a homogeneous permeability $\pPermeability$;
 and the second defining the magnet's coercive field intensity
 $\pHfieldv_{\crc}$.
The differences between them are primarily due to the 
 modelling of the coercive field intensity, 
 which is the applied field strength at which the magnet's polarity
 changes.


%
\subsection{Bulk}

\begin{figure}
\resizebox{0.95\columnwidth}{!}{%
\includegraphics{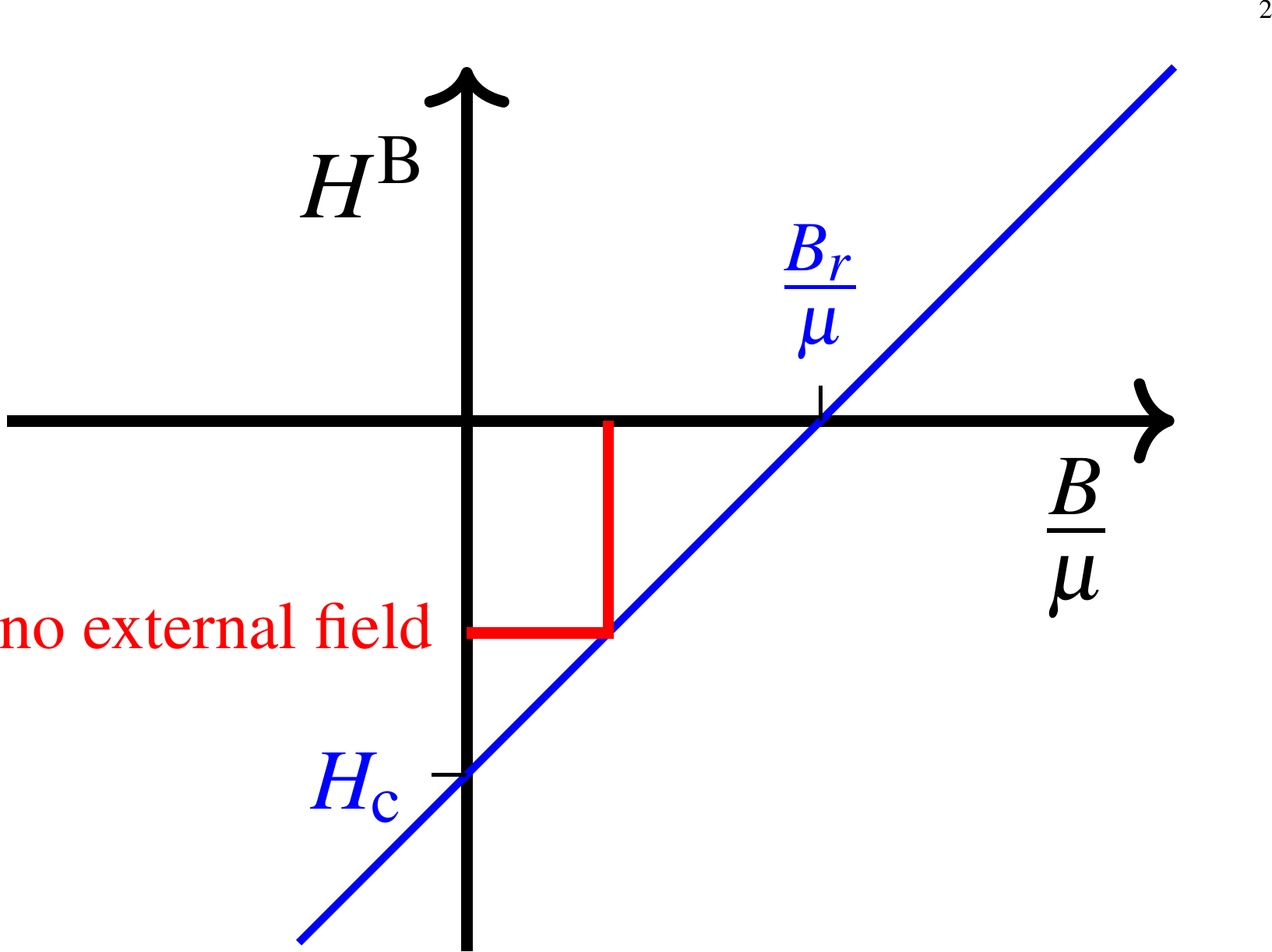}
}
\caption{The relationship
 between the component of the bulk $\pHfieldv^\bulk$ field
 and the magnetic field $\pBfieldv$
 for the bulk model (\ref{CP_perm_bulk}). 
The red lines shows the value of
  the two fields when there is no external magnetic field.}
\label{fig_HBulk_B}
\end{figure}

\begin{figure}
{\centering
\resizebox{0.95\columnwidth}{!}{%
\includegraphics{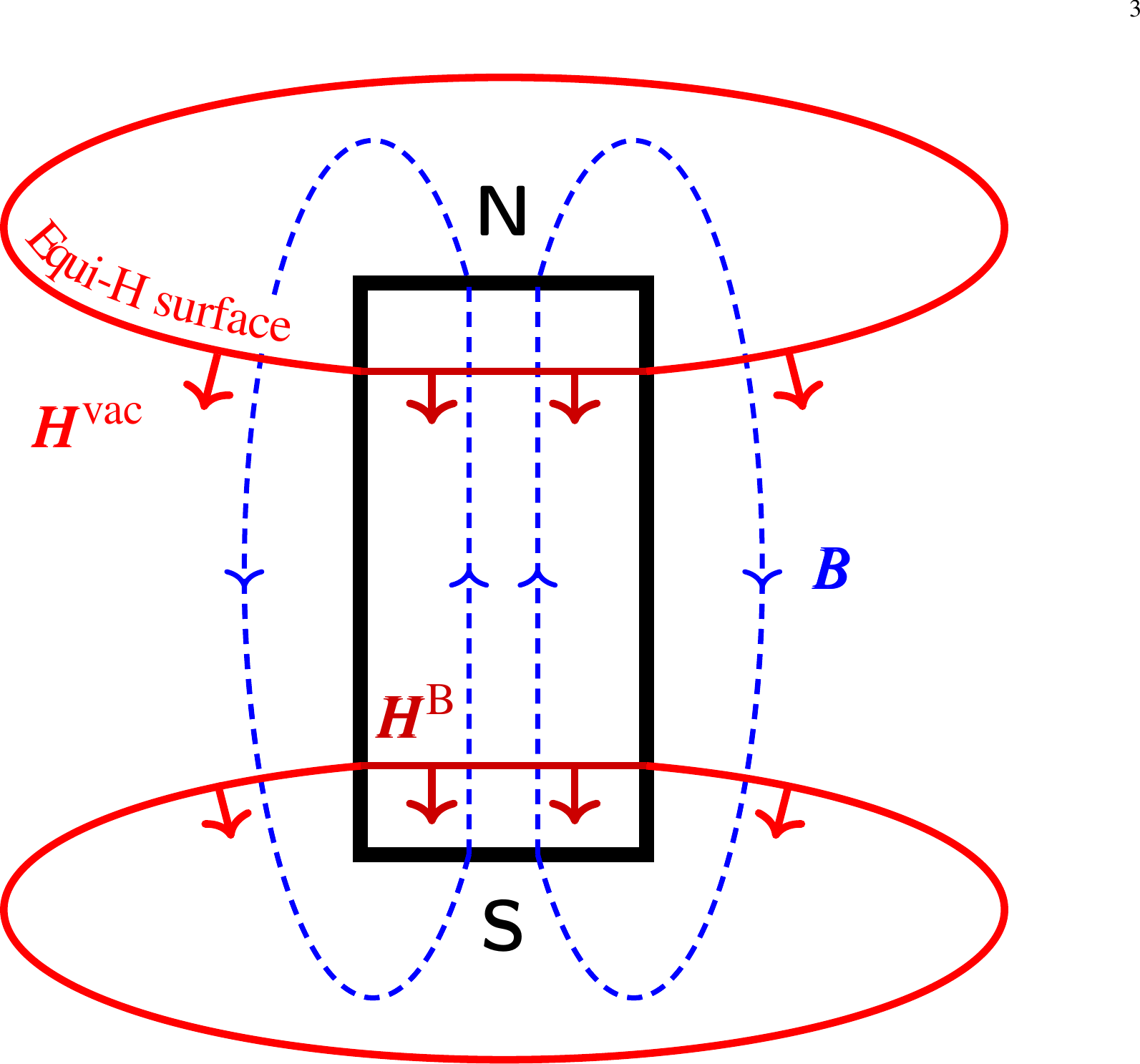}
}
}
\caption{Bulk magnetism:
A cross-section
 of the magnetic fields inside and outside
  permanent magnets.
The dashed blue lines are the $\pBfieldv$-field, 
 with the direction indicated by the blue arrows.
The red lines
 are the surfaces \emph{orthogonal} to the $\pHfieldv$-field,
 with the direction of $\pHfieldv$ indicated by the red arrows.
The vacuum $\pHfieldv$ field
 (i.e. $\pHfieldv^\vac$)
  is parallel to the $\pBfieldv$ field, 
  whereas in the medium the $\pHfieldv$ field 
  (i.e. $\pHfieldv^\bulk$)
 is anti-parallel to $\pBfieldv$. 
However the
  tangential $\pHfieldv^\bulk$ is continuous as one crosses the boundary of
  the magnet.}
\label{fig_Mag_Bulk}
\end{figure}

In the bulk model the magnetic properties of the material
 are assumed to be dispersed continuously throughout its volume, 
 along with the coercive field intensity
 $\pHfieldv_{\crc}$.
In this case, 
 the magnetic constitutive relation for $\pHfieldv=\pHfieldv^\bulk$
 is given by
\begin{align}
\pHfieldv^\bulk = \tfrac{1}{\pPermeability}\,\pBfieldv - \pHfieldv_{\crc}
\label{CP_perm_bulk}
\end{align}
where $\pPermeability>0$
 is the constant permeability of the magnetic material.
We see in fig. \ref{fig_HBulk_B} 
 that neither coercive field intensity $\pHfieldv^\bulk=\pHfieldv_{\crc}$ 
 nor the remanence field $\pBfieldv=\pBfieldv_r$
 can be achieved without an external magnetic field.
The remanence field, 
 that which remains after any applied field is removed, 
 is then 
 $\pBfieldv_r={\pPermeability}\pHfieldv_{\crc}$.
Away from other external magnetic fields
 the magnitude of $\pHfieldv^\bulk$ takes a value
 between zero and $\norm{\pHfieldv_{\crc}}$, 
 i.e.  $0<\norm{\pHfieldv^\bulk}<\norm{\pHfieldv_{\crc}}$. 
Thus from (\ref{CP_perm_bulk})
 we see that $\pHfieldv^\bulk$,
 the field witin the magnet,
 is
 in the \emph{opposite} direction to $\pBfieldv$, 
 as depicted in figure \ref{fig_Mag_Bulk}.

%
\subsection{Surface current}

\begin{figure}
{\centering
\resizebox{0.95\columnwidth}{!}{%
\includegraphics{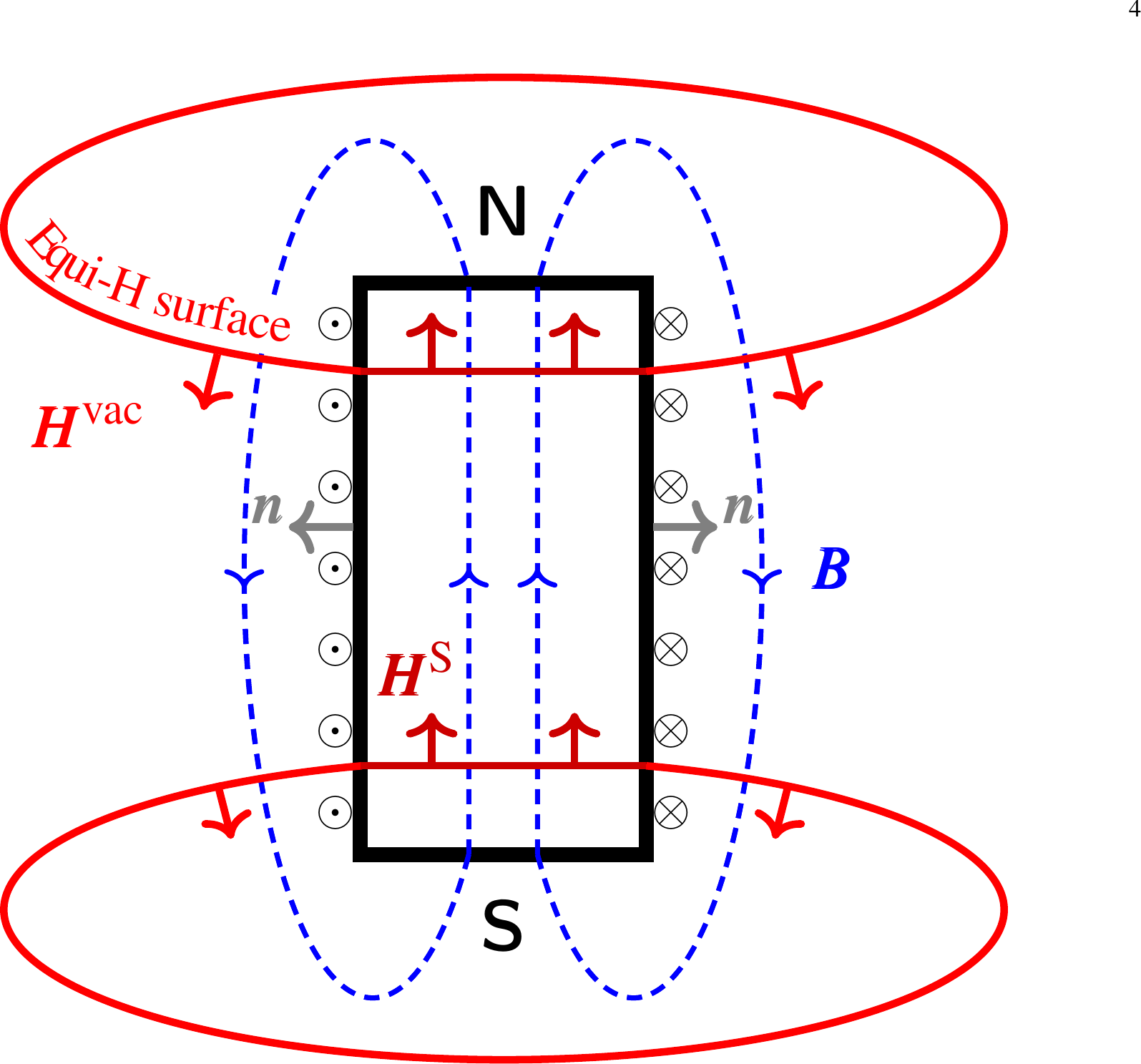}
}
}
\caption{Surface current magnetism:
A cross-section
 of the magnetic fields inside and outside
 permanent magnets. 
The dashed blue lines are the $\pBfieldv$-field, 
 with the direction indicated by the blue arrows.
The red lines 
 are the surfaces \emph{orthogonal} to $\pHfieldv$-field,
 with the direction of $\pHfieldv$ indicated by the red arrows.
The vacuum $\pHfieldv$ field
 (i.e. $\pHfieldv^\vac$)
  is parallel to the $\pBfieldv$ field, 
  just as in the medium
  where the $\pHfieldv$ field 
  (i.e. $\pHfieldv^\SC$)
  is parallel to $\pBfieldv$.
 However there is a
  discontinuity in tangential $\pHfieldv$ at the boundary of the magnet, 
  where $\pHfieldv^\vac$ and $\pHfieldv^\SC$ do not match up.
 The $\odot$ and $\otimes$ indicate the 
 direction of the surface currents,
 as they either come out of,
 or go into,
 the cross-section of the magnet represented on the page.
}
\label{fig_Mag_SC}
\end{figure}

In the surface current model the magnetic properties of the material
 are again a result of bulk properties
 dispersed continuously throughout its volume,
 but the coercive field intensity $\pHfieldv_{\crc}$
 is instead treated as being 
 due to a surface current $\pScurrentv_\bound^\SC$
 present on its exposed surfaces.
If the bulk permeability is $\pPermeability$,
 then the magnetic constitutive relation for
 $\pHfieldv=\pHfieldv^\SC$
 is given by
\begin{align}
  \pHfieldv^\SC
=
  \tfrac{1}{\pPermeability}\pBfieldv
\quad\textrm{and}\quad
  \pScurrentv_\bound^\SC 
=
  \Vec{n} \cross \pHfieldv_{\crc}
,
\label{CP_perm_surface}
\end{align}
where $\Vec{n}$
 is the outward-pointing 
 normal to the surface of the permanent magnet.
In this model $\pHfieldv^\SC$
 is in the same direction as magnetic field $\pBfieldv$, 
 as depicted on figure \ref{fig_Mag_SC}. 

\begin{figure}
\centering
\resizebox{0.95\columnwidth}{!}{%
\includegraphics{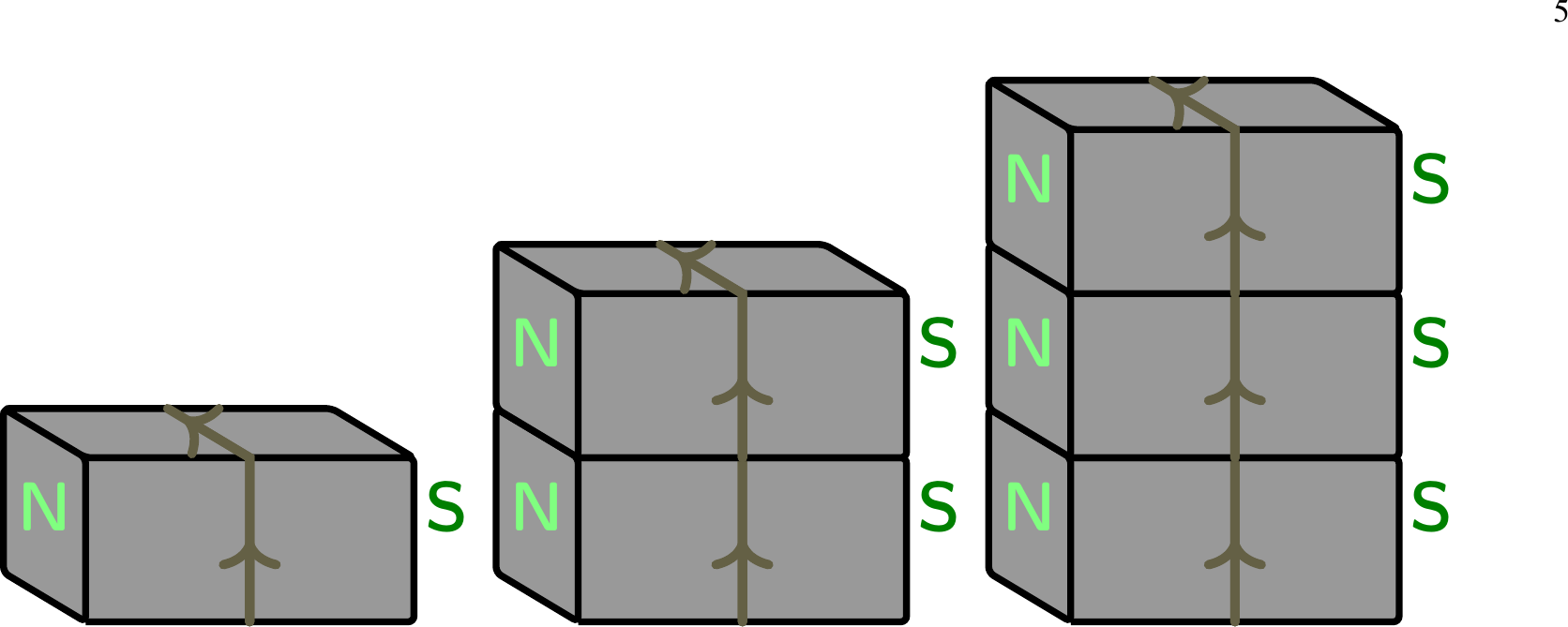}
}
\caption{Stacking magnets:
 each block has its own surface current, 
 but these cancel out on the interface where they join together,
 so the surface current of a stack is only present
 on its exterior.}
\label{fig_surface_currents}
\end{figure}

Note that 
 there is a discontinuity in $\pHfieldv^\SC$ due to the surface current, i.e.
\begin{align}
\Vec{n} \cross [\pHfieldv]=\pScurrentv_{\bound}^\SC
,
\label{CP_H_para_disc}
\end{align}
where $[\pHfieldv]=\pHfieldv^\SC-\pHfieldv^\textrm{vac}$ with $\pHfieldv^\textrm{vac}$ being the induced
magnetic field in the vacuum.
When joining two magnets (with the same orientation)
 together to create a larger magnet, 
 the surface currents on the shared interface simply cancel, 
 as shown in fig. \ref{fig_surface_currents}.


%
\subsection{Microscopic: local dipoles}

In the microscopic model the magnetic properties of the material
 are a result of bulk current-generating properties
 dispersed continuously throughout its volume,
 and a coercive field intensity $\pHfieldv_{\crc}$
 which is due to
 a collection of current-loop magnetic dipoles inside the magnet.
The dipoles generating  $\pHfieldv_{\crc}$
 average out inside the bulk,  
 and, 
 in combination with the effect of the 
 discontinuity in $\pBfieldv$ on the surface of the magnet,
 gives rise to a net surface current
 $\pScurrentv_{\bound}^\textrm{micro}$.
If the bulk permeability is $\pPermeability$,
 then the magnetic constitutive relation for
 $\pHfieldv=\pHfieldv^\textrm{micro}$
 is
\begin{align}
\pHfieldv^\textrm{micro} &= \frac{1}{\pPermeabilityVac} \pBfieldv, \\
  \pCurrentv_{\bound}^\textrm{micro}
&= 
  \left(
    \frac{1}{\pPermeabilityVac}
   -
    \frac{1}{\pPermeability}
  \right)
  \grad \cross \pBfieldv
,
\label{CP_microJ}
\\
\quad\textrm{and}\quad
 \pScurrentv_{\bound}^\textrm{micro}
&=
  \Vec{n} \cross \pHfieldv_c
 +
  \left(
    \frac{1}{\pPermeabilityVac}
   -
    \frac{1}{\pPermeability}
  \right)
  \Vec{n} \cross \pBfieldv
,
\label{CP_microS}
\end{align}
 where we have used Maxwell's microscopic equations
 along with bound currents and charges.
Note that a diagram showing the relative directions
 of $\pBfieldv$ and $\pHfieldv^\textrm{micro}$ 
 for this model would be similar to figure \ref{fig_Mag_SC}, 
 because they are straightforwardly proportional to each other,
 and therefore always in the same direction.

%
\subsection*{Comparison}

We see when comparing figs. \ref{fig_Mag_Bulk} and \ref{fig_Mag_SC}
 that the bulk and surface-current models
 give completely different predictions --
 the direction of the induced magnetic field $\pHfieldv$ in the medium
 \emph{reverses} as we switch between models. 
This stark difference means that if only 
 we could measure $\pHfieldv$ in the bulk,
 we could tell which model is the physically correct one.
However, 
 since we can only measure $\EB$, 
 Maxwell's equations \eqref{eqn-MaxwellStandard}
 insufficient to enable us to distinguish between these model.

One ameliorating feature here is that 
 inside the volume of the magnet, 
 the difference in $\pHfieldv$ between the two models 
 can be bridged by a gauge contribution:
~
\begin{align}
  \pHfieldv^\bulk
&= 
  \pHfieldv^\SC - \pHfieldv_{\crc} 
~~
=
  \pHfieldv^\SC + \grad \psi
\end{align}
where $\psi$ is an inhomogeous field given by
$\psi = -y \norm{\pHfieldv_\crc}$
where $y$ is the coordinate along the axis of the magnet, 
 i.e. where $y$ is parallel to $\pHfieldv_{\crc}$.

However, 
 the surface-current model
 has a bound surface current $\pScurrentv_{\bound}^\SC$. 
Thus to make a measurement
 that determines the correct constituitve model of a permanent magnet,
 one must either measure $\pHfieldv$ inside the medium,
 or the bound surface current $\pScurrentv_{\bound}$. 
This means that 
 as already discussed in Sec. \ref{S-measurement}, 
 since we have no mechanism for measuring $\pHfieldv$,  
 we can at best only infer $\pScurrentv_{\bound}^\SC$
 from nearby free-space $\pBfieldv$ measurements,
 a process which requires us to assume some constitutive mode.

The microscopic model is different in that 
 the bulk bound current
 $\pCurrentv_{\bound}^\textrm{micro}$
 depends on the derivative of the magnetic field $\pBfieldv$; 
 but one may consider this is less practical or aesthetically pleasing. 
However, 
 the bound currents in the other two models
 are unrelated to $\pBfieldv$. 

It is straightforward to make the empirically-based argument
 that all these models
 for the constitutive relations for a permanent magnet are equally valid.
This is because
 it is not possible to distinguish between them
 by direct measurement, 
 whether of the induced magnetic excitation field $\pHfieldv$,
 the bound bulk current $\pCurrentv_{\bound}$,
 or the surface current $\pScurrentv_{\bound}$.  
The question as to which to choose is best determined by aesthetics, 
 reference to an underlying quantum model, 
 or practical considerations such as which model is easiest to work with.


%
\subsection*{Advantages and disadvantages}

When considering
 the advantages and disadvantage of the three models,
 we note that all three arise naturally from an ``averaged'' or
 homogenised view of a magnet's microscopic properties, 
 namely current loops at the atomic or molecular level. 
The choice of model therefore depends on the aspect
 of the underlying microscopic properties the researcher wishes to emphasise.

The bulk model
 does not suffer from requiring fields which are concentrated
 in an infinitely thin region at the surface of the magnet.
These are typically expressed using delta functions
 and can produce difficulties
 when considering self-fields
 or the total energy/momentum calculated from products. 
Further,
 the lack of bound currents in a bulk model
 tends to make it easier to solve for the fields in vacuum.
However,
 a disadvantage of the bulk model is that
 one does not homogenise the current loops,
 but instead the corresponding magnetic moments of the atoms or molecules.

Despite the requirement for delta functions
 to specify the surface current in the surface current model,
 it is at least quite easy to understand --
 it is straightforward to envisage how microscopic current loops 
 representing atomic or molecular properties are averaged 
 to leave just an effective surface current
 representing the permanent magnetism.
Nevertheless,
 it does retain a somewhat artificial distinction
 between the response of the magnetic material
 (expressed as a permeability $\mu$) and the permanent magnetism.

Lastly, 
 the main strength of the microsopic (local dipole) model 
 is that it corresponds to the principle
 that a medium is really the vacuum and
 that any responses to an external field
 are (and should be)
 represented by bound currents. 
Further, 
 it does not require us to define the excitation field $\pHfieldv$, 
 although a side effect is that the constitutive relations
 for the magnet are not constant, 
 and depend on the $\pBfieldv$ field.

%
\section{Paradox? The infinite iron slab}\label{S-paradoxiis}

\begin{figure}
\resizebox{0.95\columnwidth}{!}{%
\includegraphics{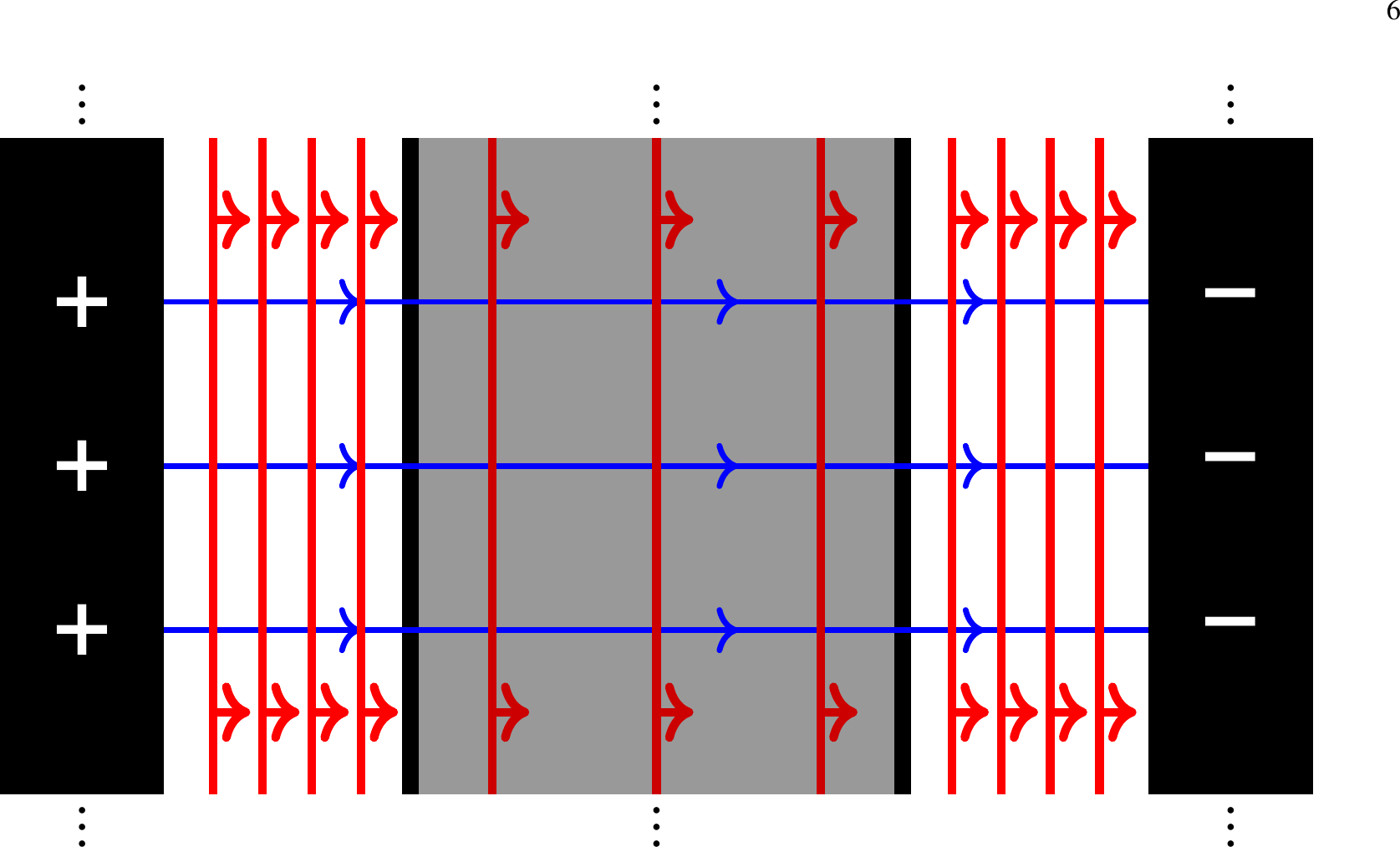}
}
\caption{
A horizontal slice through a vertically extended
 (infinite) slab of dielectric
 with an applied electric field.
 The blue lines are the
  continuous $\pDfieldv$ field, while the red lines are lines of constant
  $\pEfieldv$. Since $\pDfieldv=\pPermittivity\pEfieldv$
 and $\pPermittivity=\pPermittivityVac$ in the vacuum and
  $\pPermittivity>\pPermittivityVac$ in the medium,
  there is a reduced density of
  $\pEfieldv$ lines in the medium.}
\label{fig_Compare_Slabs_ED}
\end{figure}

\begin{figure}
\resizebox{0.95\columnwidth}{!}{%
\includegraphics{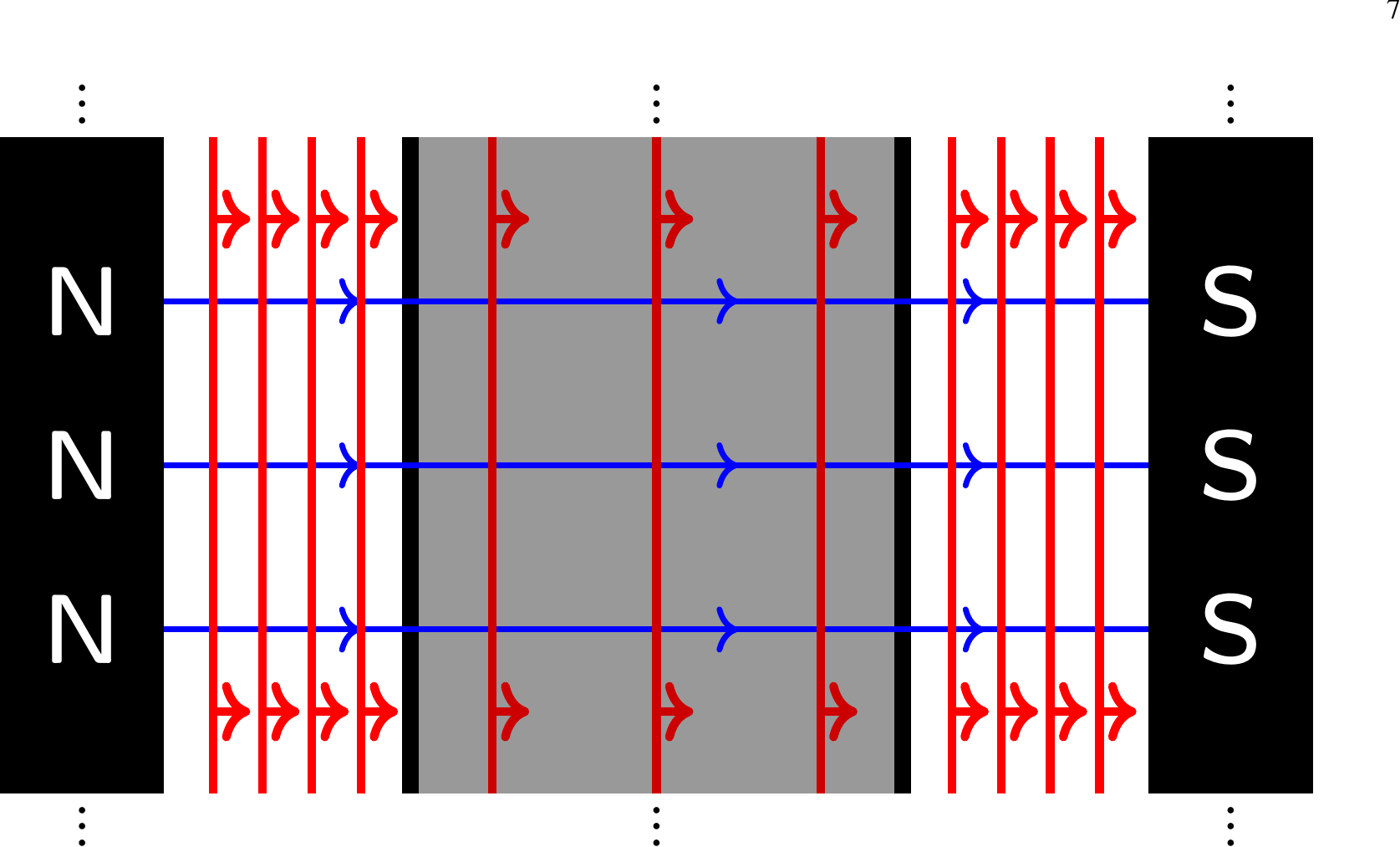}
}
\caption{
A horizontal slice through a vertically extended 
 (infinite) slab of ferromagnetic material
 with an applied magnetic field.
The blue lines are the
  continuous $\pBfieldv$ field, while the red lines are lines of constant
  $\pHfieldv$. Since $\pBfieldv=\pPermeability\pHfieldv$
  and $\pPermeability=\pPermeabilityVac$ in the vacuum and
  $\pPermeability>\pPermeabilityVac$ in the medium,
  there is a reduced density of
  $\pHfieldv$ lines in the medium.}
\label{fig_Compare_Slabs_BH}
\end{figure}

Even though our core argument has been sufficiently demonstrated above, 
 in order to further emphasize our point 
 we can compare and contrast the cases of 
 (a) a slab of dielectric medium
  between two charged plates, 
  as per fig. \ref{fig_Compare_Slabs_ED};
 and 
 (b) a ferromagnetic medium 
  with a zero coercivity
  between two magnetic poles 
  as per fig. \ref{fig_Compare_Slabs_BH}.
We see that at this diagrammatic level
 the two scenarios are completely equivalent,
 with the permittivity in fig. \ref{fig_Compare_Slabs_ED}
 acting to reduce the 
 electric field $\pEfieldv$ in the dielectric; 
 and the permeability in fig. \ref{fig_Compare_Slabs_BH}
 likewise acting to reduce the magnetic excitation $\pHfieldv$ in the medium.

\begin{figure}
\resizebox{0.95\columnwidth}{!}{%
\includegraphics{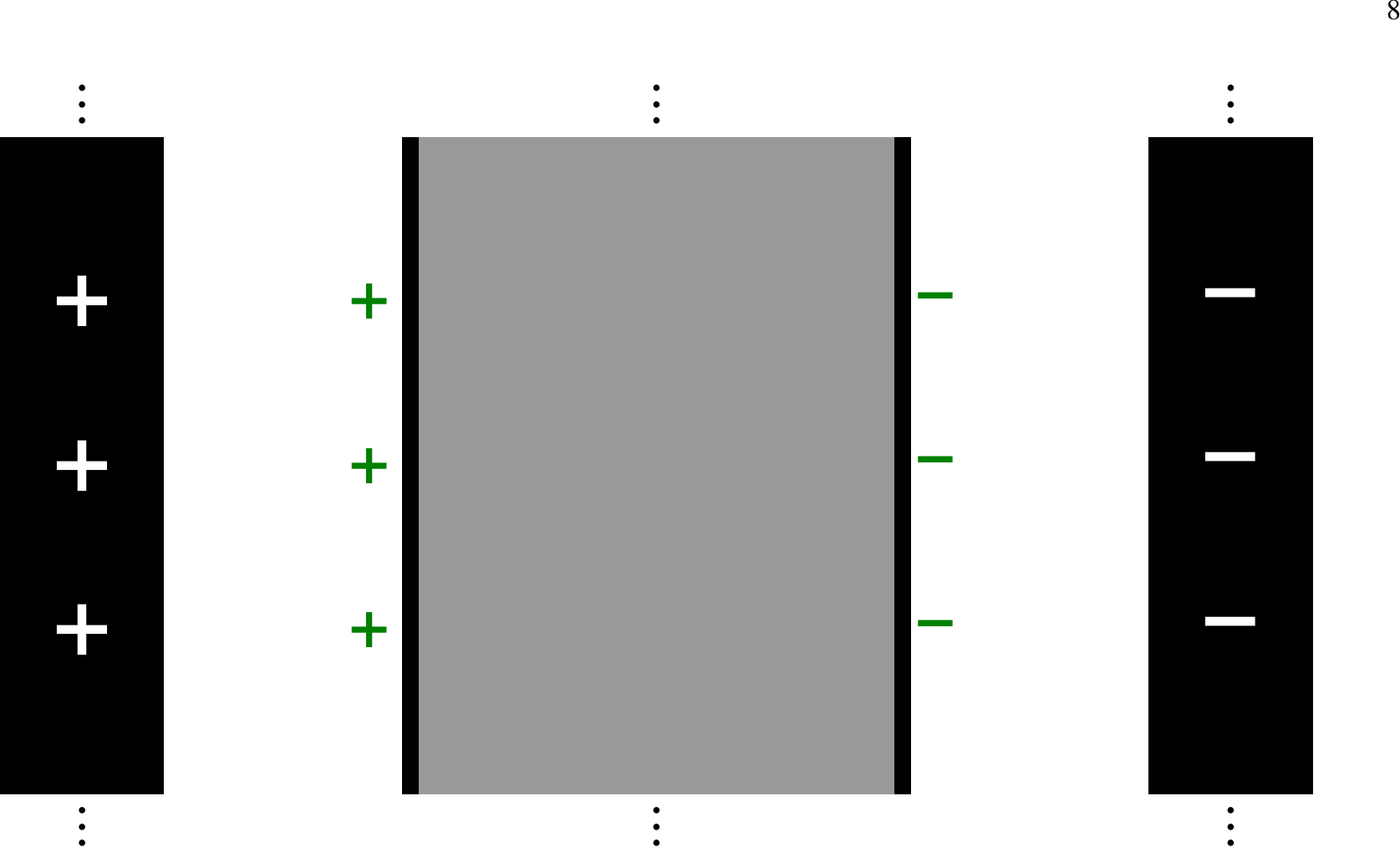}
}
\caption{Bound surface charges:
In this 
 horizontal slice through an infinitely extended dielectric slab (grey)
 in an applied electric field.
Although there are many dipoles distributed inside the dielectric,
 their constituent charges all cancel out
 except on the left and right surfaces.
Thus 
 the discontinuity in $\pEfieldv$ can be explained in terms of a
  bound surface charge distribution on the slab (green).
}
\label{fig_bounded_charge}
\end{figure}

If the slabs are finite in extent, 
 our constitutive models all behave reasonably, 
 but if we extend the slabs out towards infinity,
 an inconsistency arises when we try to explain
 the magnetisation
 in terms of \emph{bound} electric currents. 
Of course, 
 in the dielectric slab case
 the description is straightforward.
Fig. \ref{fig_bounded_charge} simply indicates that
 a bound surface dipole charge distribution
 $\pCharged_{\bound} = \grad \cdot \pEfieldv$
 is induced on the dielectric slab.

The magnetic version of this 
 has two possible interpretations.
As discussed above, 
 and given the absence of any native magnetic sources,
 Maxwell's equations \eqref{eqn-MaxwellStandard}
 insist that the magnetisation instead  
 arises from a bound current, 
 which might be either 
 dispersed throughout the volume with $\pCurrentv_{\bound}$ 
 being given by \eqref{CP_def_BoundrJ}, 
 or be present on the surface with $\pScurrentv_{\bound}$
 being given by \eqref{CP_H_para_disc}.

\begin{figure}
\resizebox{0.95\columnwidth}{!}{%
\includegraphics{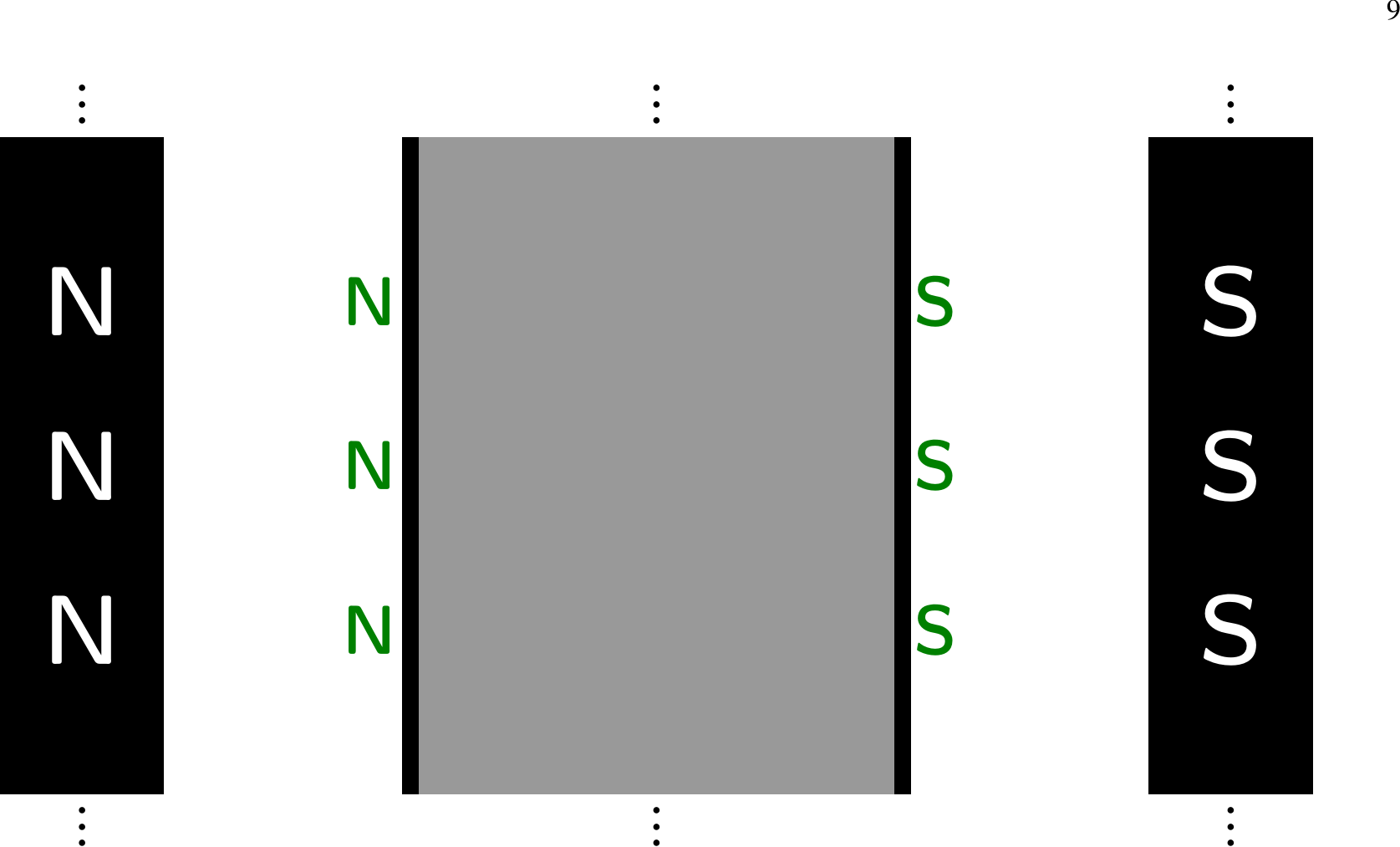}
}
\caption{Bound surface magnetic monopoles (!):
This shows a 
 horizontal slice
 through an infinitely extended slab of magnetizable material (grey),
 in an applied magnetic field.
There are induced magnetic dipoles distributed inside the material,
 and if such a thing as magnetic monopoles existed, 
 we might say that their constituent magnetic charges all cancel out
 except on the left and right surfaces.
However, 
 although the discontinuity in $\pHfieldv$ experienced
 as you cross into (or out of) the magnet
 \emph{might} be explained in terms of a 
 bound surface magnetic monopole distribution (green),
 such a situation is utterly incompatible 
 with our constitutive model of magnetization.}
\label{fig_bounded_mag}
\end{figure}

For magnetization due to bound currents 
 dispersed throughout the volume, 
 the net effect as seen externally
 is that
 the discontinuity in $\pHfieldv$
 is generated by
 a surface magnetic monopole distribution, 
 as shown on fig. \ref{fig_bounded_mag}; 
 which is similar to fig. \ref{fig_bounded_charge}.
However, 
 this is utterly incompatible 
 with our monopole-free constitutive model of magnetization, 
 and so as a consequence 
 both the description and the diagram fail
 to adequately represent the physical situation.

\begin{figure}
\resizebox{0.95\columnwidth}{!}{%
\includegraphics{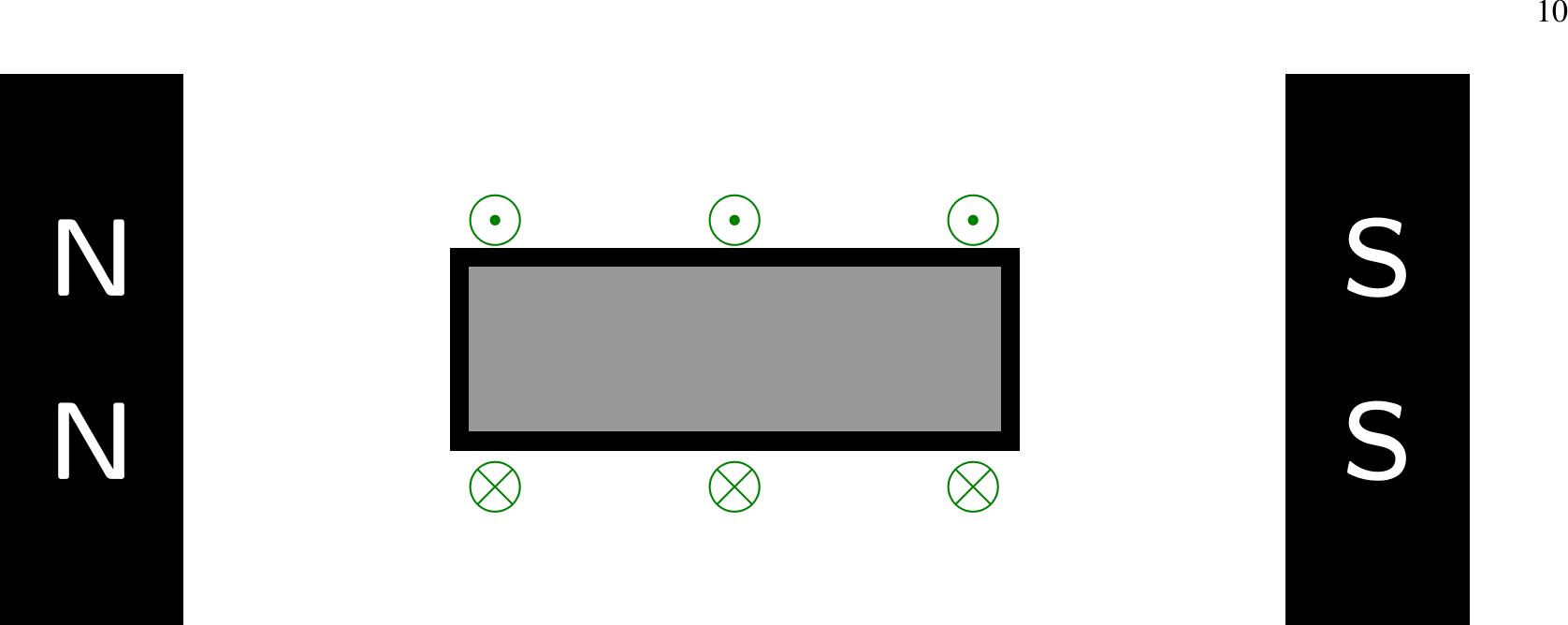}
}
\caption{Bound surface currents:
A
 horizontal slice 
 through a \emph{finite} slab of ferromagnetic material (grey)
 in an applied magnetic field.
Here
 the discontinuity in $\pHfieldv$ experienced
 as you cross into (or out of) the slab
 can be explained in terms of a
 bound surface current distribution
 whose direction
 is indicated by $\odot$ \& $\otimes$.
However, 
 if we were to stack up these finite slabs, 
 as in fig. \ref{fig_surface_currents},
 since
 the bound surface current $\pScurrentv_{\bound}$ 
 only appears on the top and bottom boundaries
 in the infinitely extended case, 
 where would any
 bound surface current
 be located?}
\label{fig_bounded_magSC}
\end{figure}

For
 magnetization due to bound surface currents, 
 the situation also not straightforward.
A depiction in the style of 
 fig. \ref{fig_bounded_charge} 
 is at least possible for the case of a \emph{finite} magnet,
 and in fig. \ref{fig_bounded_magSC} we can see 
 that the surface currents required to represent the magnetization
 can be straightforwardly placed on the top and bottom
 of the magnetic block.
However, 
 if we were to stack such blocks up one after another
 so as 
 to (eventually) form an infinite slab,
 the surface currents would cancel on the interfaces, 
 leaving only those on the exterior
 (see fig. \ref{fig_surface_currents}).
These would then be pushed further and further away as blocks were added,
 and so eventually disappear off to infinity.
There would be nowhere on the resulting figure
 where we could depict the required non-zero surface currents,
 despite them being a necessary part of the constitutive properties.

The two attempts made here 
 to consistently model the constitutive properties
 of an infinite slab of magnetized material
 have failed.
Of course, 
 although an infinite slab is not a realistic physical situation, 
 the considerations here are relevant to ``1D'' calculations
 in which a 3D system is considered unchanging in the other two.

%
\section{Conclusion}\label{S-conclusion}

In this work we have discussed how the excitation fields $\DH$ 
 can be (a) seen as gauge fields
 for the current
 (Sec. \ref{S-Maxwell}), 
 that 
 (b) there is no known way of directly measuring them 
 (Sec. \ref{S-measurement}),
 and that 
 (c) they can be given radically different values depending on
 the chosen constitutive model 
 (Sec. \ref{S-magnets}).
Further, 
 the anomalies in the interpretation
 of the standard constitutive models for magnetism, 
 as emphasized by Sec. \ref{S-paradoxiis}, 
 shows how attempts to consider $\DH$ 
 as true physical fields like $\EB$
 founder on the chosen details of the constitutive model.
This situation is particularly clear
 in the case of standard Maxwellian magnetism,
 because unlike electric effects which have their own charges,
 magnetism does not.
The subsequent necessity of describing magnetic sources 
 in terms of currents then leads directly 
 to the inconsistencies in the constitutive models.

This lack of either measurability 
 or constitutive uniqueness for the excitation field $\pHfieldv$
 in these specific magnet examples
 provides reasons
 to consider both excitation fields $\DH$
 as merely gauge fields for the current, 
 rather than quantities with an independent physical existence.
This complements the
 mathematical discussion referring to derivatives or 
 topological considerations \cite{Gratus-KM-2017nocharge},
 which also insist on the non-uniqueness of $\DH$.
The combination results in compelling arguments
 that demote $\DH$ to a gauge field status.

%
\acknowledgments

JG and PK are grateful for the support provided by
 STFC (the Cockcroft Institute ST/P002056/1)
 and EPSRC (Alpha-X project EP/N028694/1).


%

\end{document}